\documentclass[preprint]{elsarticle}
\usepackage{fullpage}
\usepackage{algorithmic}
\usepackage{mathtools}

\title{Generalization of elementary cellular automata to a higher dimension family including the BML traffic model}
\journal{Physica A}
\begin{document}

\begin{frontmatter}
\author{Daniel L. Lu}
\ead{dllu@cmu.edu, dllu@dllu.net}
\address{Carnegie Mellon University,
Robotics Institute,
5000 Forbes Ave,
Pittsburgh, PA 15213}
\begin{abstract}
    A general family of $D$-dimensional, $K$-state cellular automata is proposed where the update rule is sequentially applied in each dimension. This includes the Biham--Middleton--Levine traffic model, which is a 2D cellular automaton with 3 states. Using computer simulations, we discover new properties of intermediate states for the BML model. We present some new 2D, 3-state cellular automata belonging to this family with application to percolation, annealing, biological membranes, and more. Many of these models exhibit sharp phase transitions, self organization, and interesting patterns.
\end{abstract}
\begin{keyword}
Cellular automata; Transport phenomena; Phase transitions; Self-organization; BML model
\end{keyword}
\end{frontmatter}

\section{Introduction}
Cellular automata are useful in a variety of problems related to statistical mechanics, traffic flow, and so on. The simplest sort of cellular automaton is the \textit{elementary cellular automaton} which is one-dimensional with two states. These have interesting behaviour such as chaos and Turing-completeness, and are described extensively in Wolfram's pioneering paper \cite{wolfram} as well as later works \cite{nks, survey}.

The Biham--Middleton--Levine (BML) traffic model was first proposed in 1992 to study traffic flow \cite{bml}, as a 2D analog of the elementary Rule 184. It consists of a rectangular lattice, with periodic boundary conditions, where each site may be empty, contain a red car, or a blue car. On each time step: all red cars synchronously attempt to move one step east if the site is empty; then all blue cars synchronously attempt to move one step south if the site is empty. For some parameter $p\in [0,1]$, the model is initialized by assigning to each site either a red car with probability $p/2$, or a blue car with probability $p/2$, or empty space otherwise. The model is interesting because, for small values of $p$ it self organizes into a free flow where cars move freely without ever stopping, whereas for large $p$ it converges to a global jam. It was initially believed that there is a sharp transition between these two phases, but in 2005 a stable intermediate phase was discovered for lattices of coprime dimensions \cite{raissa5}, and then shown in 2008 to exist for square lattices \cite{raissa8}.

The BML model's simplicity and interesting behavior has inspired research that tweak every aspect of the model \cite{chowd}, such as dimensionality \cite{chowd,hku}, lattice geometry \cite{hex}, boundary conditions \cite{chowd, nonorientable, bmlr}, the update rule \cite{chowd, hku, bmlr, randomturn, overpasses, lanechange, slowtostart}, initialization \cite{chowd, accidents, initialization}, and so on. Many other cellular automaton models were proposed for modelling traffic \cite{chowd}.

Although the BML traffic model was described as an ``analog'' of Rule 184, no attempt was made to formulate a rigorous description of the analogy, or a general method of finding such analogs of other elementary rules \cite{bml}. We show that other elementary cellular automata can be generalized into two or more dimensions with arbitrarily many states, by sequentially applying the same rule in each dimension. Historically, work on two-dimensional cellular automata have mostly focused on the von Neumann or Moore neighborhood, resulting in a large neighborhood. By applying the same rule in each dimension sequentially, we only have a neighborhood of size 3 regardless of dimensionality. In this paper, we describe an extension of Wolfram's naming scheme for elementary cellular automata to higher dimensions and states; then we describe a new discovery in the BML traffic model; finally we describe and analyze a variety of different cellular automata with interesting behaviors and resembling various physical systems.

\section{Theory}
In elementary cellular automata, which have one dimension and two states, the update rule is represented as a single lookup table, which stores the state of the next cell as a function of itself and its two neighbors. The contents of the lookup table, when interpreted as a binary integer, is called the Wolfram Code \cite{wolfram}. For example, Rule 184 is a model of one-dimensional traffic flow where there is one type of car that attempts to move right if an empty site exists.

\begin{center}
\begin{tabular}{c|c|c|c|c|c|c|c}
    111 & 110 & 101 & 100 & 011 & 010 & 001 & 000\\ \hline
    1 & 0 & 1 & 1 & 1 & 0 & 0 & 0
\end{tabular}
\end{center}

Likewise, for two-dimensional, cellular automata with three states, the next cell's state can be stored as a function of itself and its neighbors. However, the von Neumann neighborhood has five sites (including the center), yielding a lookup table of size $3^5 = 243$. This is unwieldy and expensive. The Moore neighborhood has nine sites and is even bigger. Due to the large size of neighborhoods as a function of dimensionality, many two-dimensional cellular automata, such as the well-known Conway's Game of Life, simply sum up the values of their neighbors (an approach known as \textit{totalistic} cellular automata \cite{wolfram}).

For 2D cellular automata, our approach sequentially applies two updates, each of which only considers three sites (yielding a lookup table of $3^3 = 27$). That is, we alternatingly apply an update to each row, and then each column, and so on. For example, consider the BML traffic model. Let state 0 be the empty space, 1 be the car species whose turn it is to move, and 2 be the car species whose turn it is not to move. The BML update rule can be explained as ``cars of type 1 move to the next cell if possible, then turn into cars of type 2; cars of type 2 turn into cars of type 1''. Then, the table is:

\begin{center}
\begin{tabular}{c|c|c|c|c|c|c|c|c}
    222 & 221 & 220 & 212 & 211 & 210 & 202 & 201 & 200\\ \hline
    1 & 1 & 1 & 2 & 2 & 0 & 0 & 0 & 0
\end{tabular}

\begin{tabular}{c|c|c|c|c|c|c|c|c}
    122 & 121 & 120 & 112 & 111 & 110 & 102 & 101 & 100\\ \hline
    1 & 1 & 1 & 2 & 2 & 0 & 2 & 2 & 2
\end{tabular}

\begin{tabular}{c|c|c|c|c|c|c|c|c}
    022 & 021 & 020 & 012 & 011 & 010 & 002 & 001 & 000\\ \hline
    1 & 1 & 1 & 2 & 2 & 0 & 0 & 0 & 0
\end{tabular}
\end{center}

The rulestring of the BML traffic model, in base 3, is therefore 111220000111220222111220000. In base 10, it is the more concise 3922832263383. This generalizes trivially to higher dimensions and higher number of states. For $D$ dimensions and $K$ states, the length of the rulestring, written in base $K$, is of length $K^3$. The total number of rules is therefore $K^{(K^3)}$. For the rest of the paper we will focus on $D=2$, $K=3$.

The family of cellular automata we have described here is a special case of a broader class of cellular automata with \textit{cyclic} rules. Some of the assumptions in elementary cellular automata regarding symmetry and equivalence no longer hold when the rules are applied in different dimensions in sequence. For example, in elementary cellular automata, flipping all the ones and zeros makes little difference but for the BML rule above, if we change all the ones to twos and twos to ones, an entirely different model is created (see Section \ref{sandflow}).

By formulating the rulestring of such models as an integer, we can explore other interesting rules by randomly selecting one. In the following sections we discuss the BML model as well as other models with different rules. The vast majority of rules, when initialized randomly like the BML traffic model, appear to be entirely random. However, although we are sure many of these rules have complex and interesting properties such as universal computing, analysis of such rules is very difficult and we will instead focus on the ones exhibiting a visually obvious structure.

\section{Intermediate phases in the BML traffic model with extreme mobility}
\begin{figure}
    \centering
    \includegraphics[width=3cm]{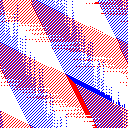}
    \includegraphics[width=3cm]{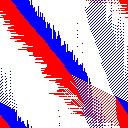}
    \includegraphics[width=3cm]{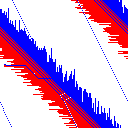}
    \caption{Two realizations of BML traffic model on a $128\times 128$ grid, with $p=0.36$, after $10^8$ iterations. From left to right: mobility $v\approx 0.91$, $v\approx 0.27$, $v\approx 0.02$.}\label{bmlex}
\end{figure}

\begin{figure}
    \centering
    \includegraphics[width=5cm]{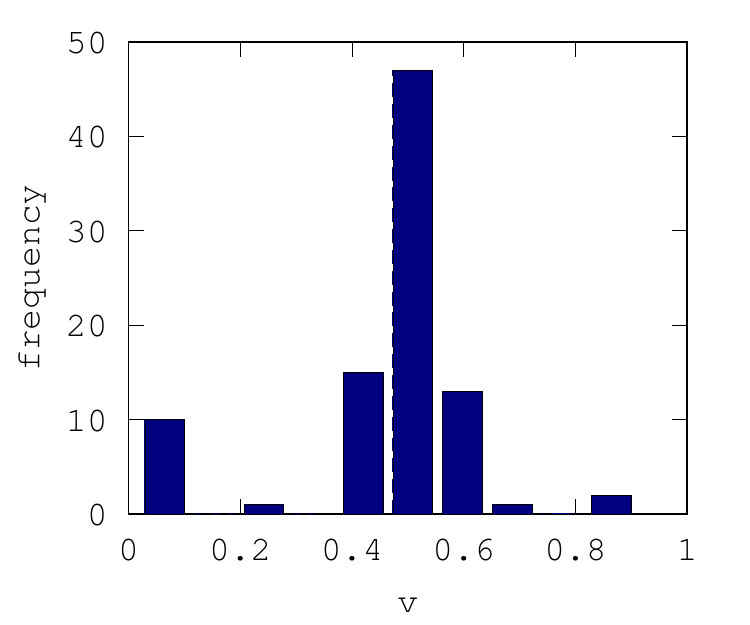}
    \caption{Histogram of mobility, $v$, for the 87 out of 2628 simulations of the BML traffic model on a $128\times 128$ grid with $p=0.36$ which did not result in a global jam.}\label{bmlhist}
\end{figure}

Our formulation of the model yields an efficient implementation on modern commodity hardware, as is discussed in Section \ref{imp}. While experimenting, we discovered the existence of very rare phases. In the BML traffic model, \textit{mobility} is the mean speed of cars -- equivalently, the ratio of cell births to cell population. We report the existence of periodic intermediate phases with very low mobility $v<0.1$ as well as very high mobility $v>0.9$. In previous studies demonstrating the existence of such phases \cite{raissa5, raissa8}, it was thought they have a mobility of around $0.5$. Our results with extreme mobility levels suggest that stable intermediate phases may exist regardless of mobility level. Some examples are shown in Figure \ref{bmlex}. Out of 2628 simulations of the BML traffic model on a $128\times 128$ grid with $p=0.36$, 2539 resulted in a global jam. None exhibited the so-called ``disordered intermediate'' state described in \cite{raissa8}. Due to the low number of non-jamming results, is unclear what exactly the distribution of intermediate phases with respect to mobility is --- a histogram is shown in Figure \ref{bmlhist}.

\section{Percolation}\label{sandflow}
A percolation cellular automaton is obtained by switching the 1 and 2 in the rulestring for the BML traffic model. In decimal, the rule number is 7469071910973. The intuitive explanation is: there are are two species of particles, where red ones alternatingly attempt to move downwards and then rightwards, and blue ones stay still. This can be thought of as another analog of Rule 184. This can be used to model the percolation of fluids or granular particles as they flow around obstacles or are deposited onto arbitrary surfaces; or even human pedestrians as they attempt to walk diagonally downwards and rightwards while avoiding obstacles.

\begin{figure}
    \centering
    \includegraphics[width=3cm]{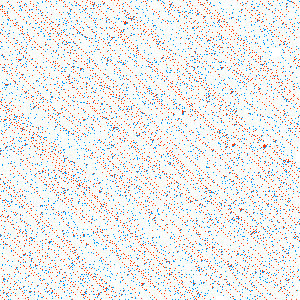}
    \includegraphics[width=3cm]{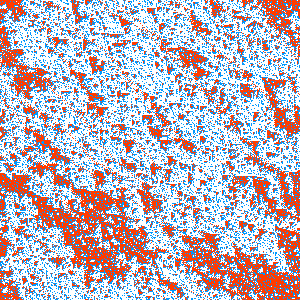}
    \includegraphics[width=3cm]{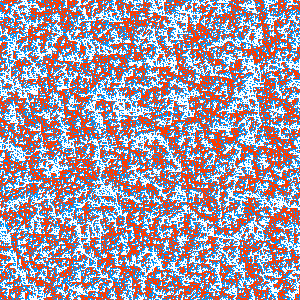}
    \caption{Rule 7469071910973: Percolation model on a $300\times 300$ grid, initialized in the same way as the BML traffic model, with $p=0.1, 0.5, 0.7$ respectively, steady state.}\label{sandfig}
\end{figure}

\begin{figure}
    \centering
    \includegraphics[width=8cm]{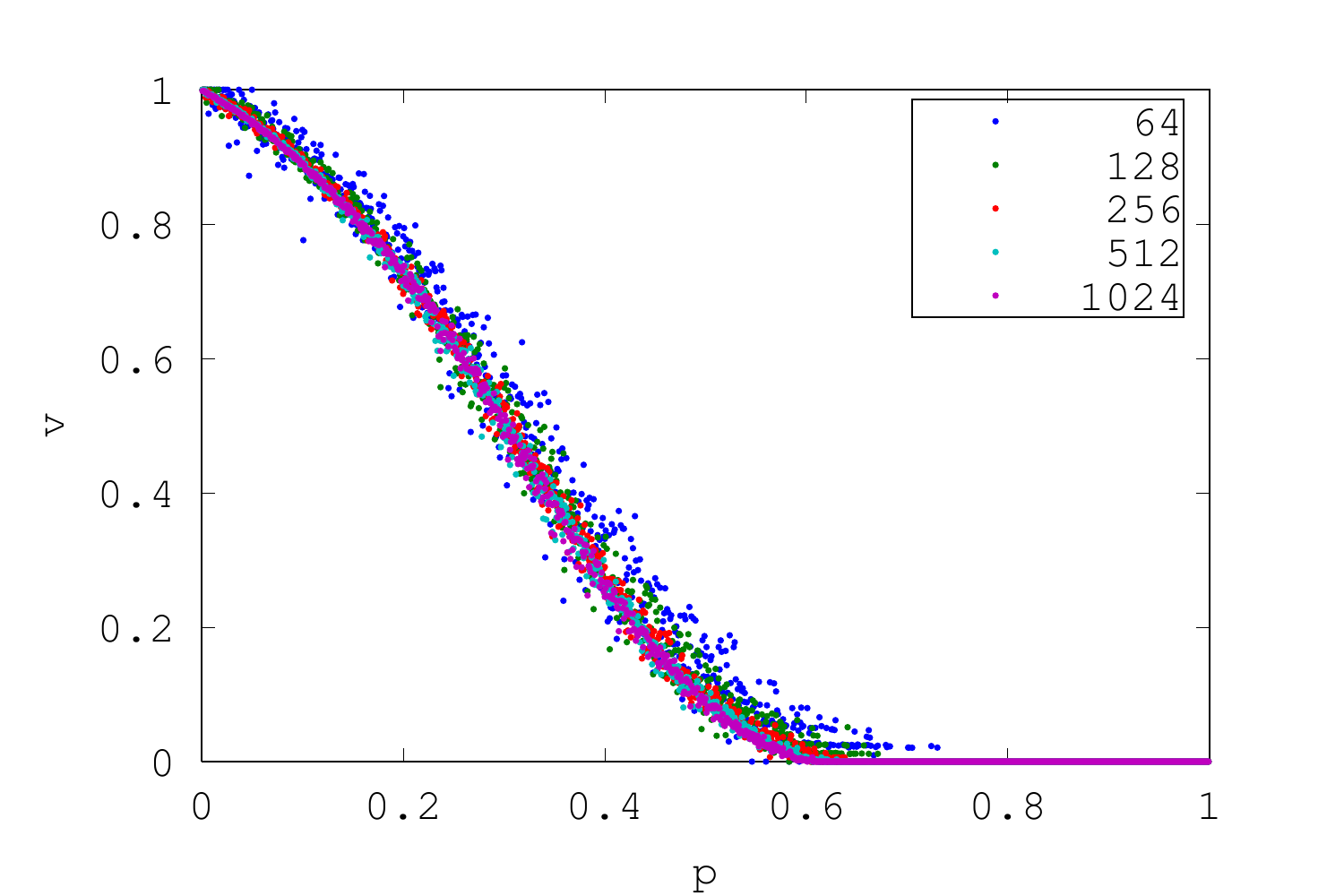}
    \caption{Graph of mobility, $v$, with respect to density $p$ for Rule 7469071910973, on square lattices of size 64, 128, 256, 512, and 1024, after $10^5$ iterations.}\label{sandv}
\end{figure}

We perform some experiments where the state is initialized in the same way as the BML traffic model, for which results are shown in Figure \ref{sandfig}. Like the BML traffic model, the red sand particles self-organize into streams that avoid obstacles when density is low; otherwise, they reach total jams. Unlike the BML traffic model, there does not seem to be a sharp transition between phases. Even when density is low, some particles can be trapped in small local jams. A plot is shown in Figure \ref{sandv}. There appears to be no dependence on lattice size. A simple calculation inspired by mean field theory provides a loose upper bound on $v$: At speed $v$, the ratio of the number of static particles (both red and blue) to the total space is $\frac{(1-v)p}{2} + \frac{p}{2}$ static particles. The mean velocity $v$ is the probability that a red particle is not next to a static particle. Assuming red particles to be uniformly distributed yields the self-consistency equation:
\begin{align}
    v = 1 - \frac{(1-v)p}{2} - \frac{p}{2} \rightarrow v = 2\frac{p-1}{p-2}
\end{align}
This calculation overestimates $v$, because it assumes the mobile particles are isotropically distributed whereas in fact only blue particles (obstacles) are; the red particles instead form streams or clump up into jams. The formation of these clumps or local jams is complex and possibly chaotic, and an analytical description is difficult. In fact, when observing the simulation for $p=0.5$, such jams often accumulate transiently at different locations before dissolving over the next iterations, only to converge far later.

Percolation theory gives another upper bound. From Figure \ref{sandv}, it appears that there is a threshold at around $p\approx0.6$ beyond which the model almost surely results in a global jam. For large lattice sizes, percolation theory states that when the density of obstacles exceeds $1-p_c$, where $p_c$ is the \textit{percolation threshold} \cite{perc, perc2}, then the lattice is no longer connected. Since the density of obstacles in our model is $\frac{p}{2}$, this implies there is no path across the lattice for any red particle when $p > 2(1-p_c)$. From past results $p_c = 0.593$, meaning our system will almost surely jam for $p>0.815$.

\section{Annealing}
Here we present a simple model demonstrating behavior similar to annealing and bearing some superficial resemblance to the square lattice Ising model. It is Rule 2828173986213, which has the following ternary representation:

\begin{center}
\begin{tabular}{c|c|c|c|c|c|c|c|c}
    222 & 221 & 220 & 212 & 211 & 210 & 202 & 201 & 200\\ \hline
     1  &  0  &  1  &  0  &  0  &  0  &  1  &  0  &  1
\end{tabular}

\begin{tabular}{c|c|c|c|c|c|c|c|c}
    122 & 121 & 120 & 112 & 111 & 110 & 102 & 101 & 100\\ \hline
     0  &  0  &  0  &  0  &  2  &  2  &  0  &  2  &  2
\end{tabular}

\begin{tabular}{c|c|c|c|c|c|c|c|c}
    022 & 021 & 020 & 012 & 011 & 010 & 002 & 001 & 000\\ \hline
     1  &  0  &  1  &  0  &  2  &  2  &  1  &  2  &  0
\end{tabular}
\end{center}

\begin{figure}
    \centering
    \includegraphics[width=3cm]{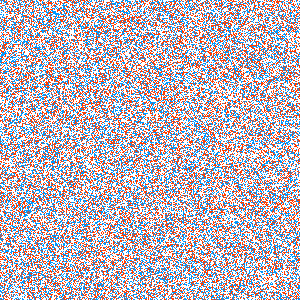}
    \includegraphics[width=3cm]{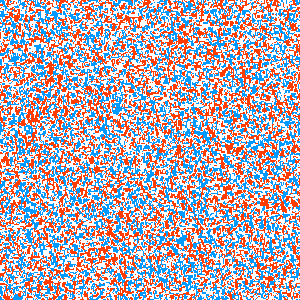}
    \includegraphics[width=3cm]{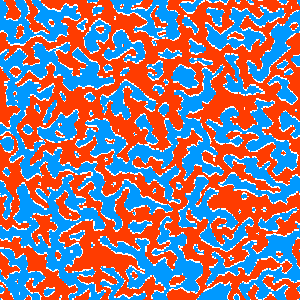}
    \includegraphics[width=3cm]{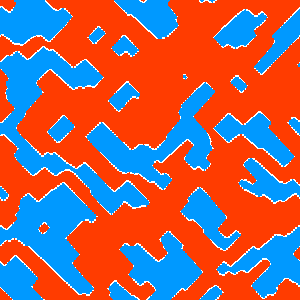}
    \includegraphics[width=3cm]{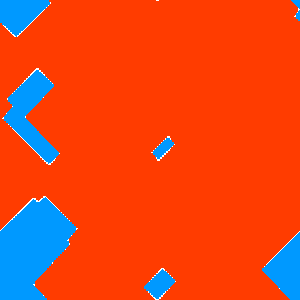}
    \caption{Rule 2828173986213: Simple annealing model on a $300\times 300$ grid, initialized in the same way as the BML traffic model, with $p=0.5$. Shown here from left to right is the same realization after timesteps $t=0, 1, 10, 100, 1090$ respectively.}\label{annealfig}
\end{figure}

Put simply, if a site's neighborhood contains only cells of one colour, the site assumes that colour. Otherwise, the site becomes empty. When the model is seeded with some initial distribution of red and blue cells (same as in the BML model), these seeds aggressively expand to cover sites in their vicinities, and soon the lattice contains some red and blue regions separated by thin borders of empty cells. These meta-stable regions have boundaries closely approximating orthogonal polygons whose sides are approximately $45^\circ$ to the horizontal. A side of such a polygon can move in the direction of its normal with speed inversely proportional to its length. As shorter sides move rapidly and join with parallel sides to be longer and stabler, the complexity of each of these regions is reduced over time and the model appears to be annealing (Figure \ref{annealfig}).

The reason why the polygonal sides move with speed inversely proportional to length is because they are not in fact \textit{exactly} oriented $45^\circ$ to the horizontal. Instead, the side's slope is often off by one unit. The one ``flaw'' in the side propagates along the side back and forth, at a constant speed; at the end of each orbit of the flaw, the side effectively moves one unit. Since the period of the flaw's orbit is proportional to the length of the side, the side advances at a speed inversely proportional to the length.

Ultimately, the model converges to either red or blue, or some simple intermediate phase (e.g. half red, half blue and separated by parallel boundaries), independent of $p$. There does not appear to be any surprising phase change behavior for this model.

\section{Membrane-like models}
In the previous section, somewhat nontrivial behavior along boundaries separating different phases was briefly discussed. More complex boundaries can form between more complex phases.

For some rulestrings, when the state is initialized in the same way as the BML traffic model, the cellular automaton rapidly partitions itself into two or more different phases separated by a membrane-like frontier. Like biological membranes, this can move in time, allow particles to cross it, merge with other membranes, and so on. A consequence is that the membranes are often transient or meta-stable. Depending on the initial choice of $p$, the model may ultimately converge to one of the different phases.

\begin{figure}
    \centering
    \includegraphics[height=3cm]{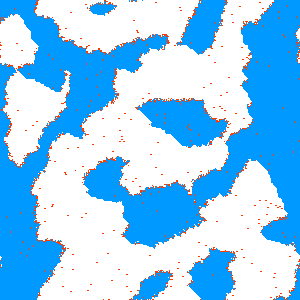}
    \caption{Rule 152690720768. Typical transient state on a $300\times 300$ lattice with $p=0.66$.}\label{152690720768}
\end{figure}

\begin{figure}
    \centering
    \includegraphics[width=8cm]{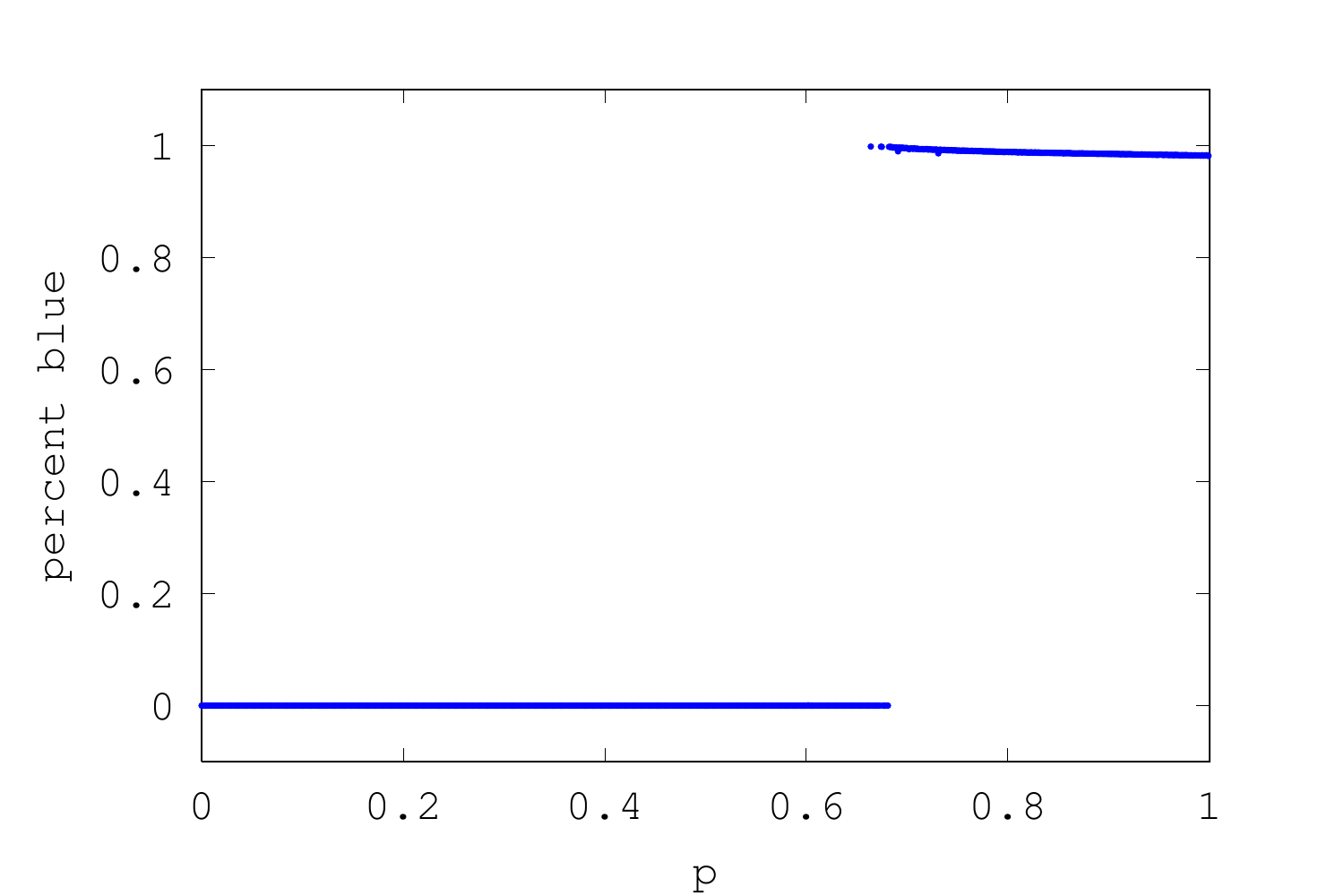}
    \caption{Rule 152690720768. A plot of the percentage of lattice occupied by blue cells after $10^5$ iterations, on a $1024\times 1024$ lattice. The transition appears to occur at around $p=0.68$.}\label{membraneplot}
\end{figure}

One such model is Rule 152690720768, shown in Figure \ref{152690720768}. This model was found by random rulestring generation and then visual inspection. Its microscopic behavior is too complex to be studied in detail here, so instead we perform experiments to determine its sensitivity to $p$. The model appears to have a sharp transition around $p=0.68$, above which the blue phase dominates and below which the white phase dominates (Figure \ref{membraneplot}). Intriguingly, there are red cells sparsely interspersed throughout both phases.

\begin{figure}
    \centering
    \includegraphics[width=3cm]{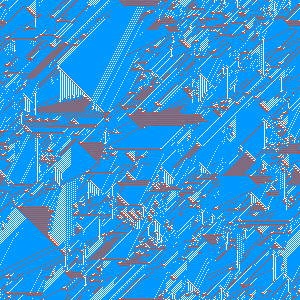}
    \includegraphics[width=3cm]{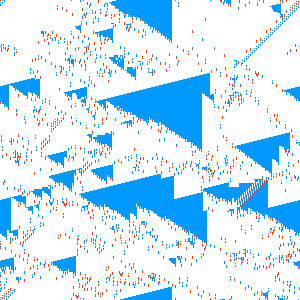}
    \includegraphics[width=3cm]{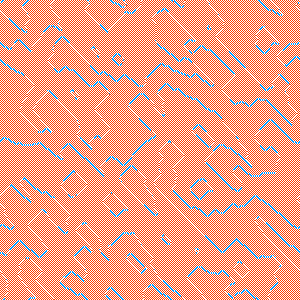}
    \includegraphics[width=3cm]{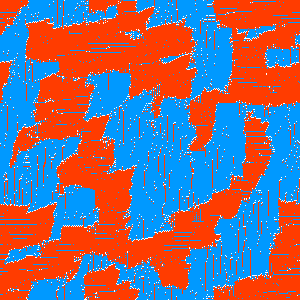}\\
    \includegraphics[width=3cm]{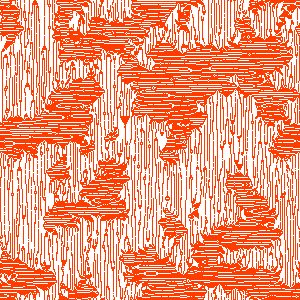}
    \includegraphics[width=3cm]{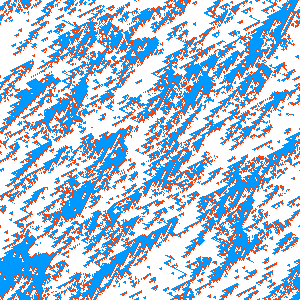}
    \includegraphics[width=3cm]{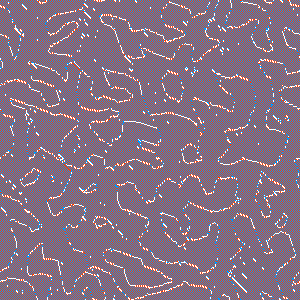}
    \includegraphics[width=3cm]{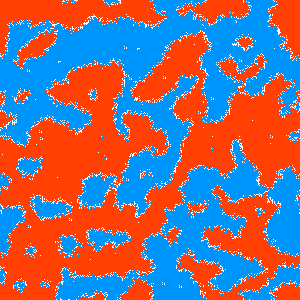}
    \caption{Rules 5882493049933, 1811177701721, 5340268068864, 3894972317834, 2403954491737, 400697024, 3245024084244, 456351711232, simulated on a $300\times 300$ lattice.}\label{interesting}
\end{figure}

\section{Implementation}\label{imp}
Study of cellular automata such as the BML traffic model is severely restricted by available computing power. For a moderately sized $128\times 128$ lattice, each time step requires on the order of $10^4$ site updates. Supposing the model takes up to $10^6$ time steps to converge, this already takes several minutes on a typical CPU assuming a typical $10^8$ updates per second. Past studies have used purpose-built hardware such MIT's CAM8 architecture \cite{raissa5}, for which development has unfortunately stopped in 2001 due to a disk crash caused by ``mindless jerks'' \cite{cam8}. Even then, it took a whole month for the simulations to run. Later researchers have implemented the BML traffic model on a graphics processing unit (GPU) \cite{nonorientable}, which is suitable for the task due to the embarrassingly parallel nature of the model.

Our implementation also uses the GPU to process several rows of the lattice in parallel. Like \cite{nonorientable}, our implementation uses the CUDA technology; the graphics card used in our experiments is a single NVIDIA GeForce GTX 970, which has 1664 CUDA cores \cite{gtx970}.

As a minor note, in the implementation it is important to use a good random number generator for large simulations. Many common programming languages have a built-in pseudorandom number generator that has a period of $2^{32}$, far too small for generating $10^5$ simulations of size more than $10^4$ each. Our implementation uses the Mersenne Twister with a period of $2^{19937}-1$.

\section{Conclusions}
The main contribution of this paper is the proposal of a family of cellular automata operating on a neighborhood of size 3 for arbitrary $D$ dimensions and $K$ states, which operates by sequentially applying the same update rule in each dimension. This family includes the widely-studied BML traffic model, for which we present new results regarding the mobility of intermediate states. Furthermore, we have briefly discussed the properties of some other cellular automata within this family, and their possible application to various problems including percolation and annealing. The advantages of using a cellular automaton for such problems include the simplicity of the model and ease of implementation.

Future work includes a more thorough exploration of such models (Figure \ref{interesting}), as well as models with different numbers of dimensions and states. In particular, although this paper talks about the $D=2, K=3$ case, the simpler case of $D=2, K=2$ is yet to be explored. For this, there are only 256 rules, the same number as elementary cellular automata. For implementation, a possible idea for future optimization is the Gigantic Lookup Table optimization \cite{glut}, which groups adjacent sites into a block that is updated all at once.

Our implementation is available open source at \texttt{https://bitbucket.org/dllu/bml-cuda/}.

An interactive visualization for alternative rules is available at \texttt{http://www.dllu.net/bml/}.

\end{document}